\documentclass[prb,twocolumn,showpacs,superscriptaddress,aps]{revtex4-1}
\usepackage{graphicx,amssymb,amsmath,psfrag,color}
\usepackage[colorlinks=true,citecolor=blue,linkcolor=blue,urlcolor=blue]{hyperref}
\begin{document}
\definecolor{darkgreen}{rgb}{0,0.5,0}
\newcommand{\be}{\begin{equation}}
\newcommand{\ee}{\end{equation}}
\newcommand{\jav}[1]{\textcolor{red}{#1}}

\title{Gauge field entanglement of Kitaev's honeycomb model}

\author{Bal\'azs D\'ora}
\email{dora@eik.bme.hu}
\affiliation{Department of Theoretical Physics and MTA-BME Lend\"{u}let Spintronics 
Research Group (PROSPIN), Budapest University of Technology and Economics, 1521 Budapest, Hungary}
\author{Roderich Moessner}
\affiliation{Max-Planck-Institut f\"ur Physik komplexer Systeme, 01187 Dresden, Germany}

\date{\today}

\begin{abstract}
A spin fractionalizes into matter and gauge fermions in Kitaev's spin liquid on the honeycomb lattice.
This follows from a Jordan-Wigner mapping to fermions, allowing for the construction of minimal entropy ground state wavefunction on the cylinder. 
We use this to calculate the entanglement entropy by choosing several distinct partitionings.
First, by partitioning an infinite cylinder into two, the  $-\ln 2$ topological entanglement entropy is reconfirmed.
Second, the reduced density matrix of the gauge sector on the full cylinder is obtained after tracing out the 
matter degrees of freedom.
This allows for evaluating the gauge entanglement Hamiltonian, which contains infinitely long range correlations along the symmetry axis of the cylinder.
The matter-gauge entanglement entropy is $(N_y-1)\ln 2$ with $N_y$ the circumference of the cylinder.
Third, the rules for calculating the gauge sector entanglement of any partition are determined. 
Rather small correctly chosen gauge partitions can still account for the topological 
entanglement entropy
in spite of long-range correlations in the gauge entanglement Hamiltonian.


\end{abstract}

\maketitle

\section{Introduction.}

Fractionalization is a  ubiquitous phenomenon in a variety of systems, when strong correlation and quantum fluctuations, usually with a healthy dose of 
frustration, give way
to exotic states of matter. 
The best known instances include spin-charge separation in Luttinger liquids\cite{giamarchi}, fractional 
quantum Hall states\cite{martin} spin ice materials\cite{castelnovo}, spin liquids\cite{zhourmp}, 
just to mention 
a few.
Their elementary excitation cannot be constructed as simple combinations of its elementary constituents. 
Parallel to these developments, topological states of matter enjoy a great deal of interest due to their potential in  fault-tolerant quantum computation\cite{laumann}.
These are usually characterized by non-local, topological order, unlike conventional phases of matter such as e.g. ferromagnets.
Their robust  topological degeneracy and possibly  non-abelian statistics make them ideal systems to realize quantum computers protected from environmental noise.

The peculiar interplay of fractionalization and topological order is beautifully demonstrated in Kitaev's spin liquid on the honeycomb lattice\cite{kitaev2006}.
This system of interacting spins, introduced in Eq. \eqref{kitham}, does not order down to zero temperature,  with the  
spin pair correlations  
are identically zero beyond nearest neighbor separation \cite{baskaran,knolle}. 
This stems from the fact that a spin 
fractionalizes into matter and gauge degrees of freedom and has a 
description involving a  matter sector  coupled to a $Z_2$ gauge field, the latter being also responsible for the topological order.

Topological order is notoriously difficult to detect not only experimentally\cite{plumb,banerjee} but also theoretically due to the lack of local order parameter.
Often it is easier to detect in terms of what is absent, rather than what is present. 
To circumvent this problem, the entanglement entropy of topologically ordered phases has been shown to contain a topological contribution\cite{hamma,levinwen,kitaevpreskill},
which is uniquely sensitive to the underlying topological structure.
This can be separated from the more conventional entanglement 
contributions\cite{eisert} by carefully partitioning the system so that these are subtracted off.

Here we investigate the question that to  what extent only one constituent of a fractionalized topological system reflects the existence of topological order. 
For that reason, we study Kitaev's honeycomb model on a cylinder, which is the "closest" geometry to a torus and has recently been investigated  extensively to address the physics of the topological entanglement entropy\cite{balents,depenbrock}.
In particular, what is the entanglement entropy between gauge and matter fermions on a cylinder with open edges? 
What is the entanglement spectrum of the gauge field over the whole cylinder?
How is this related to the spectrum of the parent Hamiltonian?
Can only one (i.e. the gauge) sector account for the topological entanglement entropy?

This we investigate by tracing out the matter degrees of freedom from the ground state density matrix, thus constructing the reduced density matrix for the gauge degrees of freedom.
In an analogous case, one dimensional fermions fractionalize into spin and charge degrees of freedom, which, in the absence of spin-orbit coupling, are decoupled.
For the Kitaev model, since the ground state lives in the zero flux sector\cite{kitaev2006}, it is tempting to think that all this involves is a unique static gauge configurations, therefore
the gauge and matter fermions are completely disentangled.
In addition, while much is known about the matter field entanglement 
properties of Kitaev's honeycomb lattice\cite{hongyao,mandal2016,pachos}, the gauge field contribution has received significantly less 
attention. We also fill this gap and elaborate on the topological and non-topological entanglement entropy solely from the gauge sector.

\section{Kitaev's honeycomb model}

\begin{figure}[h!]
\centering
\includegraphics[width=6.7cm]{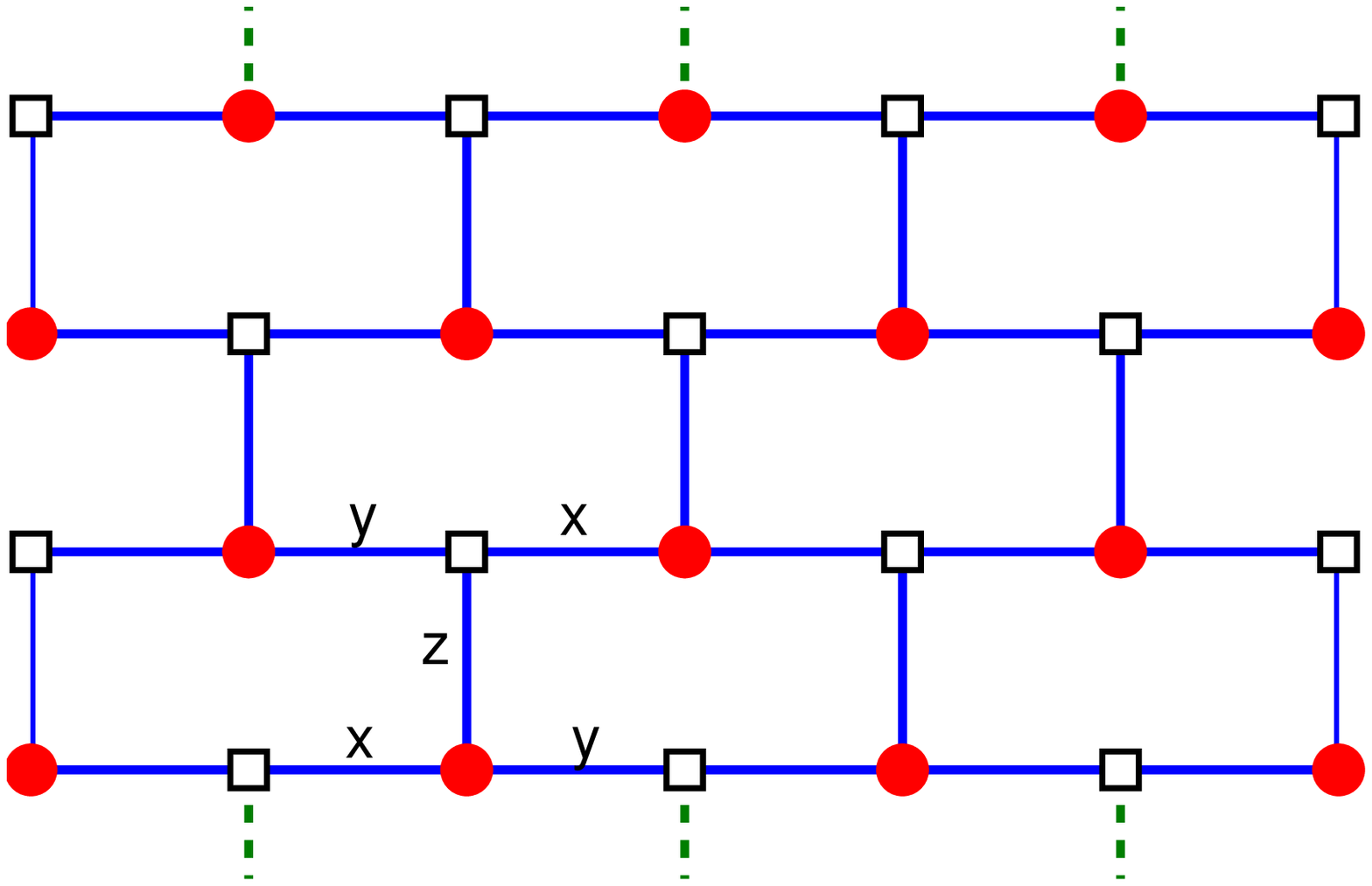}
\includegraphics[width=16mm]{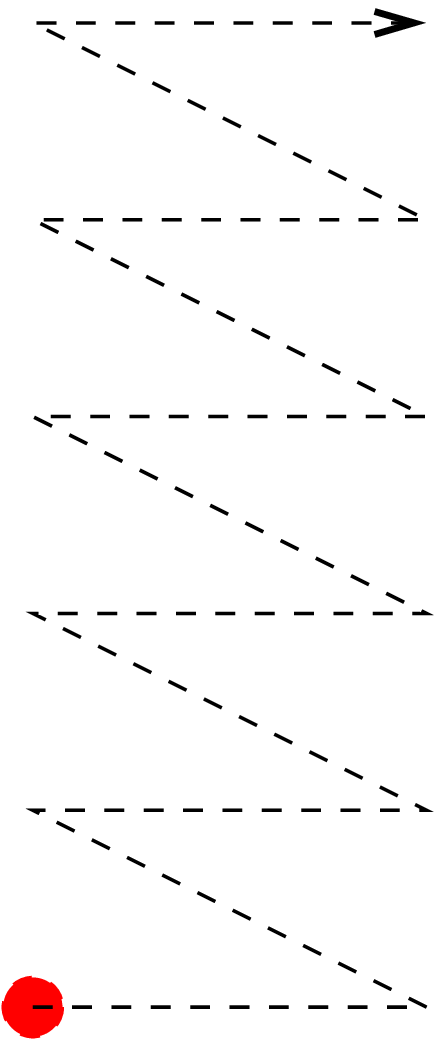}
\caption{A segment of the brick wall lattice, equivalent to the honeycomb lattice, is shown in the left panel for $N_x=7$ and $N_y=4$.
 The bottom row is connected to the
top row (green dashed lines) for periodic boundary conditions in the vertical direction, thus forming a cylinder.
The vertical bonds denote $J_z$ coupling, while the $J_{x,y}$ couplings around a given site are distributed anti-clockwise (i.e. $J_x$, $J_y$ and $J_z$).
The spins are represented by fermions after a Jordan-Wigner transformation, and the two sublattices are denoted by filled red circles and empty black squares. 
The spins are ordered from left to right, bottom to top.
 The right panel depicts the Jordan-Wigner strings and the ordering of the spins.
\label{bwjw}}
\end{figure}

The Hamiltonian is defined on the honeycomb lattice, similarly to graphene, as
\begin{gather}
H=-\sum_{\gamma=x,y,z}J_\gamma\sum_{\gamma-bonds}\sigma^\gamma_i\sigma^\gamma_j,
\label{kitham}
\end{gather}
where $\gamma=(x,y,z)$ stands for the three type of bonds, connecting sites $i$ and $j$, as explained and visualized in Fig. \ref{bwjw}, where the
topologically equivalent brick wall lattice is used. The very same lattice geometry has been used also to identify topological order for the 
Haldane model with hardcore bosons\cite{cincio}.
Instead of using the Majorana fermion representation of the spins\cite{kitaev2006} which leads unphysical states in the Hilbert space requiring a projection operation,
we employ a Jordan-Wigner transformation\cite{chen} \emph{without} redundant degrees of freedom.
This is given by
\begin{gather}
\sigma^z_i=2c^+_i c_i-1,\hspace*{6mm} \sigma^+_i=\left[\prod_{j<i}\sigma^z_j\right]c^+_i,
\label{jwtrafo}
\end{gather}
where $j<i$ is taken from the particular ordering of the spins, moving from left to right, bottom to top, see Fig. \ref{bwjw}.
The $\prod_{j<i}\sigma^z_j$ expression is referred to as the Jordan-Wigner string.
Following Ref. \onlinecite{chen},
we introduce
Majorana fermions on the two sublattices of the brick wall lattice, denoted by \emph{f}illed circles and \emph{e}mtpy squares, as
\begin{subequations}
\begin{gather}
B_f=(c-c^+)_f/i,\hspace*{3mm} A_f=(c+c^+)_f,\\
A_e=(c-c^+)_e/i,\hspace*{3mm} B_e=(c+c^+)_e,
\end{gather}
\label{majoranas}
\end{subequations}
and $A_{e,f}^2=B_{e,f}^2=1$.
After some algebraic manipulation, the Hamiltonian becomes
\begin{gather}
H=-iJ_x\sum_{x-bonds}A_fA_e-iJ_y\sum_{y-bonds}A_fA_e-\nonumber\\
-iJ_z\sum_{z-bonds}\alpha_rA_fA_e,
\label{hmajorana}
\end{gather}
where $\alpha_r=iB_fB_e$ along a $z$-bond is a conserved quantity with $r$ the location of the $z$-bond.
It contains one Majorana fermion from each end of the bond, and represent the gauge degrees of freedom, while the $A$
Majorana fermions, which hop around the honeycomb lattice, correspond to the matter sector.
A physical spin fractionalizes into these two sectors, as follows from Eqs. \eqref{jwtrafo} and \eqref{majoranas}.
The gauge field, $\alpha_r$ is hermitian and $\alpha_r^2=1$, thus its eigenvalues are $\pm 
1$, hence the name $Z_2$ spin liquid.
For the Kitaev Hamiltonian, the flux per plaquette is a conserved quantity, 
defined by
\begin{gather}
W_p=\prod_{j \in plaquette} \sigma^{\gamma_j}_j,
\end{gather}
where $\gamma_j=(x,y,z)$ corresponds to the bond from site $j$ that is not
part of the loop around a 6 site plaquette. After a Jordan-Wigner 
transformation, it reduces to the product of the two $\alpha_r$ operators in a give plaquette. 

The matter fermion spectrum is gapless and possesses a linearly dispersing 2D Dirac cone whenever the sum of any two couplings, $J_{x,y,z}$ is bigger than the third.
Otherwise, the system is gapped.

We consider a brick wall lattice on a cylinder, i.e. with periodic boundary conditions in one 
direction, as shown in Fig. \ref{bwjw}. The number of rows, $N_y$, in this direction is thus
even. For simplicity, we fix the number of spins in the horizontal direction, $N_x$, to be odd.
There are $N_yN_x$ spins in the system with $N_xN_y/2$ $z$-bonds, while the number of plaquettes is
$(N_x-2)N_y/2$. Thus, after fixing the flux sector of the model by defining a flux for each hexagon,
we end up with $N_xN_y/2-(N_x-2)N_y/2=N_y$ conserved $\alpha_r$ operators, whose value can be chosen freely without altering the overall flux configuration. 
Therefore, on a cylinder, there are $2^{N_y}$ distinct wavefunctions corresponding to a given flux configuration.
Let us stress that other orientations of the Jordan-Wigner string and other ordering of the 
spins is also possible\cite{mandal}, but for the sake of simplicity, we pick this most
transparent configuration. It has \emph{no} accidental Majorana zero modes from its boundary or dangling Majoranas\cite{dassarma}.

\begin{figure}[h!]
\includegraphics[width=6cm]{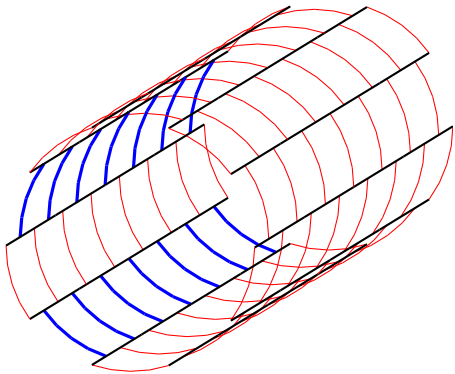}
\caption{The zero flux sector of the Kitaev model on a cylinder with circumference $N_y=10$ and $N_x=15$ is shown. The blue and red $z$-bonds stand for 1 and $-1$ 
values of the gauge fields along a given row.
Since there are 2 blue and 8 red rows,  $W_l=1$ and this configuration corresponds to the true ground state.
A cylinder with an odd number of blue rows corresponds to $W_l=-1$.
\label{bwcylinder}}
\end{figure}

\section{Ground state sector}

The ground state of the system lies in the zero flux sector\cite{lieb,kitaev2006} for large $N_y$ (for e.g. $N_y=2$, the ground state of the system of a $\pi$-flux phase on a two-leg ladder\cite{feng}). 
In the zero flux sector on the cylinder, all $z$-bonds along a given row should be identically $1$ or $-1$, 
as visualized in Fig. \ref{bwcylinder}, but 
one has the freedom to choose between these two 
configurations. 
By taking the product of the values of the row-ending $z$-bonds along 
the circumference or $N_y$ direction is again a constant of motion and corresponds to a Wilson-loop operator\cite{mandal} as
\begin{gather}
W_l=\prod_{j\in loop}\sigma^{\gamma_j}_j,
\end{gather}
the loop winds around the cylinder, the definition of $W_l$ parallels the plaquette operator and $\gamma_j$ denotes the bond that is not
part of the loop at site $j$. From Ref. \onlinecite{kitaev2006}, it equals the product of gauge variables, $\alpha_r$, along the loop as $W_l=\prod_{r\in loop}\alpha_r$.

\begin{figure}[h!]
\psfrag{x}[t][][1][0]{$N_y$}
\psfrag{y}[b][t][1][0]{$\Delta E/J_z$}
\includegraphics[width=8.5cm]{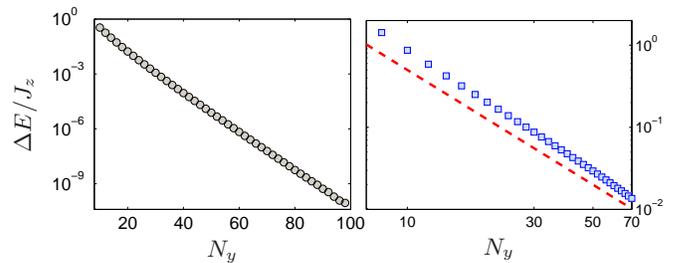}
\caption{The energy gap from numerical diagonalization of Eq. \eqref{hmajorana} between the two 
$2^{N_y-1}$ degenerate manifolds in the zero flux sector, corresponding to $W_l=\pm 1$, is plotted in the gapped ($J_x=J_y=0.4J_z$, left panel, circles) and gapless ($J_x=J_y=J_z$, right panel, squares) 
phases of the Kitaev 
model for 
$N_x=103$ on a semilog and loglog scale, respectively. The red dashed line denotes $\sim N_y^{-2}$.
\label{gsenergy}}
\end{figure}

\begin{figure}[h!]
\psfrag{x}[t][][1][0]{$J_x(=J_y)/J_z$}
\psfrag{y}[b][t][1][0]{$2E_0/|J_z|N_xN_y$}
\includegraphics[width=7cm]{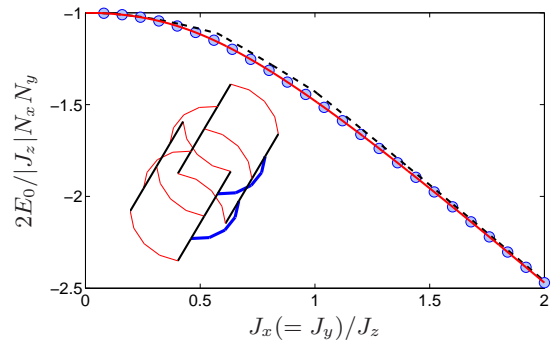}
\caption{The ground state energy, $E_0$ calculated for the $N_x=5$, $N_y=4$ cylinder  (depicted in the inset) from Lanczos diagonalization of Eq. \eqref{kitham} (circles) 
for $J_x=J_y$. For comparison,
numerical results for the zero flux sector from Eq. \eqref{hmajorana} for the ground state energy with $W_l=-1$ (red line) and the energy for 
$W_l=-1$ (black dashed line) are plotted.
Both curves are 8-fold degenerate.
\label{energynx5ny4}}
\end{figure}

On the cylinder, $N_y$ is even thus $N_y/2$ is an integer. Depending on whether it is even/odd, the ground state is in the $W_l=-1$/1 sector.
In this case, similarly to the toric code, the zero flux sector splits in energy and the true finite-size 
ground state has only a $2^{N_y-1}$-fold
degeneracy, separated by a gap from the other zero flux
states. This gap scales exponentially/in a power-law fashion with $N_y$ in the gapped/gapless phase of the Kitaev model. This is demonstrated in Fig. \ref{gsenergy} by
numerically diagonalizing Eq. \eqref{hmajorana} in the zero flux sector and calculating the ground state of the matter fermions for the two distinct $W_l=\pm 1$ configurations.
We have also confirmed the $2^{N_y-1}$ ground state degeneracy of the zero flux sector by  numerical diagonalization of Eq. \eqref{kitham}
up to 20 spins on small cylinders, similarly to Ref. \onlinecite{dassarma}, as shown in Fig. \ref{energynx5ny4}.


Each gauge field configuration in the zero flux ground state is uniquely represented by its wavefunction $|u\rangle$.
Therefore, a total ground state wavefunction, including both the matter and gauge fields, is a Schr\"odinger-cat state consisting of an equal weight superposition 
of all the $2^{N_y-1}$ gauge field configurations in the ground state of the zero flux sector. This state is fully symmetric, and in particular, 
all symmetries of the spin correlation 
functions are restored\cite{dassarma}.
This symmetric configuration is written as\cite{oshikawa,balentsprb}
\begin{gather}
|\Psi\rangle =\frac{1}{\sqrt{2^{N_y-1}}}\sum_u |u\rangle \otimes |\phi(u)\rangle,
\label{wf}
\end{gather}
with $|\phi(u)\rangle$ a matter fermion wavefunction on the background of the $u$ gauge configuration, which is that of a BCS superconductor\cite{kitaev2006,chen},
and $\langle u'|u\rangle=\langle \phi(u')|\phi(u)\rangle=\delta_{u',u}$.
The sum includes only those $2^{N_y-1}$ gauge configuration which are dictated by the parity of $N_y/2$.
This is the minimal entropy state\cite{oshikawa}, which knows about the $Z_2$ spin liquidness of the Kitaev model, as we demonstrate below.

We note and show below that both the gapped and gapless ground-states in the Kitaev model possess topologically ordered wavefunctions,
signaled by the nonzero topological entanglement entropy. However,  only the gapped Kitaev Hamiltonian (rather
than the gapless Kitaev Hamiltonian) has a well-defined topological order.

\section{Cutting the cylinder into two}

With the above symmetric ground state wavefunction without redundant degrees of freedom at our disposal,
we can extract its entanglement properties relatively straightforwardly. 
The conventional way of characterizing the topological properties of the system is the take 
an infinite cylinder\cite{balents,depenbrock} (or at least $N_x\gg N_y$) with both gauge and matter degrees of 
freedom,
partition it
into two identical regions ($A$ and $B$)
by cutting it into two identical corner-free half cylinders and trace out all (matter and gauge) degrees of freedom in the $B$ part.
This is achieved to yield
\begin{gather}
\rho_A=\textmd{Tr}_B\rho=\frac{1}{2^{N_y-1}}\sum_v|v\rangle\langle v|\otimes \rho_A^m(v),
\label{rhoa}
\end{gather}
where the $v$ gauge fields reside only on half of the cylinder and $\rho=|\Psi\rangle\langle\Psi|$ is the ground state density matrix.
It satisfies the same constraint as in Eq. \eqref{wf}, namely only the $W_l=1$ or $-1$ configurations are considered from the zero flux sector.
While the gauge field contribution is already diagonal, the matter fermion part, $\rho_A^m(v)$ in general is not, and it carries all the information about
the matter fermionic band structure, non-local correlation and possibly Abelian vs. non-Abelian phases\cite{hongyao}.
Due to the specific form of Eq. \eqref{rhoa}, its entanglement entropy factorizes\cite{swingle}
as
\begin{gather}
S=S_m+(N_y-1)\ln 2,
\end{gather}
where $S_m$ is the entanglement entropy from $\rho_A^m(v)$, while the second expression stems solely from the gauge fields.
It contains the topological entanglement entropy, $S_{topo}=-\ln 2$, as expected also from the toric code\cite{kitaev2003,oshikawa}. 
Our calculation represents one of the rare instances when this contribution can be obtained
using elementary analytical steps.

\section{Entanglement between gauge and matter}

Since the physical spin fractionalizes into gauge and matter fermions, evidenced by Eq. \eqref{majoranas}, it is an interesting question to ask how these two sectors are entangled with respect to each 
other.
Naively, since the ground state occupies the zero flux sector, one could think that there is no entanglement between these two components.
However, from the ground state density matrix,  the matter fermions are traced out, and the reduced
density matrix of the gauge field is obtained as
\begin{gather}
\rho_g=\textmd{Tr}_m\rho=\frac{1}{2^{N_y-1}}\sum_u |u\rangle\langle u|,
\label{rhog}
\end{gather}
being completely diagonal. From this,  the reduced density matrix of the gauge field has $2^{N_y-1}$ degenerate eigenvalues, having the value of $2^{1-N_y}$.
The R\'enyi entropy, calculated from  $S_\alpha=\frac{1}{1-\alpha}\ln\textmd{Tr}\rho_g^\alpha$ with $\alpha$ a positive number, is
obtained as $(N_y-1)\ln 2$, independent of $\alpha$. 
Note that, since the matter field has already been traced out, this is the total entropy without any other non-universal term, depending on the specific couplings on the lattice.
Interestingly, it has the very same structure as expected for the topological entanglement entropy on a cylinder\cite{kitaevpreskill,levinwen,balents,depenbrock,huang}, being
proportional to the circumference, and having a $-\ln 2$ constant value, as expected for minimal entropy states\cite{oshikawa}.

Eq. \eqref{rhog} allows us to construct the corresponding entanglement Hamiltonian, $H_E$, whose density matrix at an entanglement temperature $T_E=1$ 
is exactly $\rho_g$. From the gauge variables in each row, a collective spin operator is built as $S^z_j=\sum_{r\in j\textmd{th } row}\alpha_r$.
Its two extreme eigenvalues correspond to all $\alpha_r$ being 1 or $-1$ in the $j$th row. These are the only states which live in the zero flux sector.
 By projecting onto them, an effective two-level system, $\tau^z_j$, is defined in each row.
These are then used to construct the entanglement Hamiltonian as 
\begin{gather}
H_E=\lambda \prod_j \tau^z_j,
\label{gaugeham}
\end{gather} 
with $\lambda\rightarrow\pm\infty$. The sign is chosen to  ensure that the proper half of the zero flux manifold
is retained from the $2^{N_y}$ possibilities, as dictated by the parity of $N_y/2$.
Then, taking $\exp(-H_E)$ and normalizing it with the partition function yields directly $\rho_g$.

We discuss the universality\cite{chandran} of the gauge entanglement Hamiltonian by varying the entanglement temperature.
One can think of  $\rho_g$ as taking the thermal density matrix of $H_E$ at a fictitious entanglement temperature, $T_E=1$, i.e. $\rho_g\sim \exp(-H_E/T_E)$.
The entanglement temperature dependence of various quantities can be investigated by considering arbitrary $T_E$'s.
Due to the $|\lambda|\rightarrow\infty$ condition, the entanglement temperature drops out completely from $\rho_g$ and its entropy is constant, $(N_y-1)\ln 2$, for
all entanglement temperature. This is in sharp contrast to the finite temperature behaviour of the parent Hamiltonian\cite{nasu}, Eq. \eqref{kitham}, where
even for a frozen gauge field configuration, the mobile matter fermions contribute significantly to the temperature dependence of the entropy and other observables.

\section{Topological entanglement entropy in the gauge sector?}

The topological entanglement entropy, which is sensitive to the quantum dimension of the underlying model, can be determined by partitioning an infinite cylinder into two.
However, from a numerical perspective, it suffers from the drawback that it contains non-universal terms as well which can be eliminated by finite size scaling in the circumference
of the cylinder.
However, other, more elaborate methods to extract directly this topological information are famously available\cite{levinwen,kitaevpreskill}. By carefully partitioning the system and adding/subtracting the various 
entanglement
contributions, one ends up only with the topological entanglement entropy, without additional non-universal terms. 
This requires, for the present case, to consider the spatially reduced density matrix, containing both matter and gauge degrees of freedom.
Instead of carrying out this calculation, we ask the intriguing question to what extent only one component of the system, in this case only the gauge fields,
know about the topology. From Ref. \onlinecite{hongyao}, it sounds plausible to conclude that the gauge sector is responsible for the topological entanglement entropy.

\begin{figure}[h!]
\includegraphics[width=85mm]{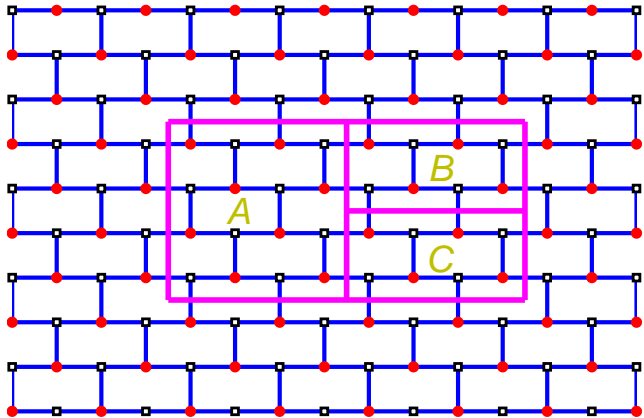}
\caption{A specific incarnation of the Kitaev-Preskill construction to obtain the topological entanglement entropy is sketched on the brick wall lattice.
\label{kptee}}
\end{figure}

Here we investigate this statement and evaluate the (topological) entanglement entropy from only the reduced density matrix of the gauge sector, Eq. \eqref{rhog}.
In order to do so, the various partitions are required to be larger than the correlation
length in the system. However, the reduced density matrix of the gauge sector contains inherently long range correlations from tracing out the matter fields:
knowing the value of the gauge field on a given $z$-bond immediately fixes all $z$-bond values in a given row, thus in some sense, infinitely long range entanglement sets in.
Thus, no region is large enough compared to the correlation length in the reduced density matrix. Nevertheless, as we argue,
as long as we consider partitions without disconnected rows, the topological entanglement entropy is obtained correctly.

Let us warm up with the entanglement of a minimal gauge "bit", a single gauge  Majorana fermion (i.e. either $B_f$ or $B_e$) with the rest.
Since $\alpha_r=iB_fB_e$ is a constant of motion, one in practice needs to focus only on this reduced Hilbert space. By doubling this local Hilbert space
artificially without introducing
any coupling between the two copies, we consider the effective Hamiltonian
\begin{gather}
H_{eff}=iB_fB_e+iB'_fB'_e
\label{twomajo}
\end{gather}
 with the $B'$ ancilla Majoranas.
The entanglement entropy between the $e$ and $f$ Majoranas is obtained\cite{pachos} after tracing out the $e$ Majoranas
 as $S=\ln 2$. Since it comes from partitioning two independent Majorana pairs into single Majoranas,
we conclude that the contribution
of cutting a single Majorana pair into two Majoranas gives $S_{majorana}=\ln\sqrt{2}$. This agrees with what is expected from the thermodynamical degeneracy of a single Majorana mode\cite{dassarma}.
From this, it follows that a minimal gauge degree of freedom is maximally entangled with the rest of the system.
This entanglement entropy  also describes  the entanglement between the two Majorana modes at the ends of a one dimensional topological superconductor\cite{laumann}.

The entanglement entropy from the gauge fields of a given spatial region is determined by following the rules:

a) each row with at least one full $z$-bond in a given partition contributes with $\ln 2$.

b) each broken $z$-bond contributes with a single Majorana and gives $\frac 12 \ln 2$ to the entropy (in accord with our discussion below Eq. \eqref{twomajo}).

c) the $x$ and $y$ bonds do not contribute explicitly to the entanglement.

By following these steps, we evaluate the topological contribution to the entanglement entropy using the Kitaev-Preskill construction, shown in Fig. \ref{kptee}.
The entanglement entropy of the various partitions is evaluated  as
\begin{subequations}
\begin{gather}
S_B=S_C=3\ln 2,\hspace*{5mm} S_{A}=S_{BC}=5\ln 2,\\ S_{AB}=S_{AC}=S_{ABC}=7\ln 2.
\end{gather}
\label{entropies}
\end{subequations}
Putting this together, we get
\begin{gather}
S_{topo}=S_A+S_B+S_C+S_{ABC}-\nonumber\\
-(S_{AB}+S_{AC}+S_{BC})=-\ln 2.
\end{gather}
This agrees with what is expected from the quantum dimension\cite{kitaev2006,hongyao,balents}, $\mathcal D=2$ of the Kitaev model, namely that $S_{topo}=-\ln\mathcal D$.

Among these, rule a) and b) are sensitive to the spatial extension of the partition in the $y$ and $x$ directions, respectively, therefore an area-law \cite{eisert} applies also
for the gauge field entanglement.
While rule a) stems from the long range entanglement in Eq. \eqref{gaugeham} and is also responsible for the topological entanglement entropy,
b) is irrelevant for the topology and is only needed to account properly for the entanglement entropy of a given partition.
For example, by halving region $A$ vertically in Fig. \ref{kptee}, two broken $z$-bonds are lost, one from the top and one from the bottom following rule b).
This reduces only the entanglement entropies, containing partition $A$, by
$\ln 2$. Since $A$ appears in four partitions, these entropy changes compensate each other.


Other partitions work equally well as long as no disconnected rows appear, the boundaries do not even need to be parallel with the bonds.
We can also treat the case of partitioning the reduced density matrix of the gauge sector on the cylinder into two identical half-cylinders by cutting it into two in the middle. 
In this case, all 
$N_y$ bonds are affected at the boundary and there are no broken $z$-bonds, so rule a) would naively give $N_y\ln 2$ entanglement. However, 
due to the constraint that only $W_L=1$ or $-1$ 
makes the ground state sector, we end up with $(N_y-1)\ln 2$ entropy. This contains the topological as well as the area part, but is free from non-universal 
terms.

Another construction, proposed by Levin and Wen\cite{levinwen} naturally uses partitions with disconnected boundaries, therefore disconnected rows
appear in certain partitions, which inevitably violate to condition of considering regions larger than the correlation length.
We stress that by using this construction, we usually get integer multiples of $\ln 2$, depending on the partitions and not zero for the topological 
entanglement entropy. This results from the fact that the gauge entanglement Hamiltonian, Eq. \eqref{gaugeham}, yields infinitely long range correlations, thus
no partition satisfies the condition of being large compared to the correlation length. This proves vital when disconnected regions are considered.
We speculate that the Kitaev-Preskill construction could be useful to extract topological information numerically for other systems\cite{balents,depenbrock} 
even for partition smaller than the correlation 
length, similarly to our case.

\section{Summary}

We have investigated the entanglement properties of Kitaev's honeycomb model (more precisely, the equivalent brick wall lattice) on a cylinder by focusing on the gauge field contribution.
By using a  Jordan-Wigner mapping to fermions without redundant degrees of freedom, 
the ground state wavefunction as a minimal entropy state was constructed without any need for projection into the physical Hilbert space. 
After partitioning it into two identical half cylinders, its entanglement entropy satisfies the "sum rule" that the contributions from the matter  and gauge sector are additive\cite{swingle}.
The gauge entanglement entropy is $(N_y-1)\ln 2$ with $N_y$ the 
circumference of the cylinder, and the $-\ln 2$ is identified as the topological entanglement entropy of a $Z_2$ spin liquid\cite{balents,depenbrock}.

The reduced density matrix of the gauge sector from the minimal entropy state is obtained after tracing out the matter degrees of freedom. 
It assumes a diagonal form with all $2^{N_y-1}$ eigenvalues being equal to $2^{1-N_y}$, consequently all gauge-matter R\'enyi entropies are $S_\alpha=(N_y-1)\ln 2$.
The gauge entanglement Hamiltonian contains infinitely long range correlations along the symmetry axis of the cylinder. 
The basic rules for calculating the gauge field entanglement entropy are determined.  We demonstrated  that even small
partition \`a la Kitaev and Preskill provide the right topological entanglement entropy in spite of long range correlations in the entanglement Hamiltonian.

\begin{acknowledgments}

We thank J. Asb\'oth, S. Mandal and F. Pollmann  for insightful discussions.
This research is supported by the National Research, Development and Innovation Office - NKFIH K108676, SNN118028 and K119442 and by the  DFG via SFB 1143.

\end{acknowledgments}

\bibliographystyle{apsrev}
\bibliography{refspinliquid}

\end{document}